\documentclass{elsart}

 \usepackage{graphicx}
\usepackage{amssymb}
\usepackage{amsmath}

\begin{document}
\begin{frontmatter}



\title{Moving contact line with balanced stress singularities}


\author{X. Y. Hu}
\author{and N. A. Adams}

\address{Lehrstuhl f\"{u}r Aerodynamik, Technische Universit\"{a}t
M\"{u}nchen
\\ 85748 Garching, Germany}
\begin{abstract}
A difficulty in the classical hydrodynamic analysis of moving
contact-line problems, associated with the no-slip wall boundary
condition resulting in an unbalanced divergence of the viscous
stresses, is reexamined with a smoothed, finite-width interface
model. The analysis in the sharp-interface limit shows that the
singularity of the viscous stress can be balanced by another
singularity of the unbalanced surface stress. The dynamic contact
angle is determined by surface tension, viscosity, contact-line
velocity and a single non-dimensional parameter reflecting the
length-scale ratio between interface width and the thickness of the
first molecule layer at the wall surface. The widely used Navier
boundary condition and Cox's hypothesis are also derived following
the same procedure by permitting finite-wall slip.
\end{abstract}
\end{frontmatter}
\section{Introduction}
Immiscible two-phase flows with moving contact lines occur in a
variety of applications, such as coating and biological processes.
The moving contact line problem, however, has for many years
remained a partially open issue. One of the problems is the validity
of the no-slip wall boundary condition, which arises with classical
hydrodynamics, where for a no-slip wall an unbalanced divergence of
the viscous stress occurs, which leads to a violation of the
contact-angle condition at a moving contact line
\cite{HuhScriven1971}\cite{Dussan1979}. There have been many
attempts to resolve the problem by modifying the boundary condition,
including the slip model \cite{Hocking1977}\cite{ZhouSheng1990}, the
interface relaxation model \cite{Shihkmuraev1997}, the diffusive
interface model \cite{Jacqmin2000} \cite{ChenJasnowVinals2000}, and
the combined molecular-dynamics and diffusive-interface model
\cite{QianWangSheng2003}. However, studies by molecular dynamics
show that even though considerable contact-line velocities can be
obtained \cite{KoplikBanavarWillemsen1988}
\cite{ThompsonRobbins1989}, the maximum shear rate is still many
orders less than taht which can violate the no-slip wall boundary
condition considerably \cite{ThompsonTroian1997}.
\section{Smoothed, finite-width interface model for moving contact line}
In this Letter, we reexamine the hydrodynamics of a
fluid/fluid/solid system with a steady moving contact line. Instead
of considering sharp interfaces directly, our analysis starts from
smoothed, finite-width interfaces. Given the continuous interfacial
free energy density with the form \cite{Jacqmin1999}  $f =
\frac{1}{2} \sigma |\nabla C|^2 + \Psi(C)$, where $C$ is a color
function, $\sigma$ is a coefficient and $\Psi(C)$ is the bulk energy
density, at the state which minimizes $\mathcal{F}=\int f dV$ the
interface reaches its equilibrium profile. In this case the surface
stress $\Pi_{ij}$ in a two dimensional Cartesian coordinate system
is given by
\begin{equation}\label{surface-stress}
\Pi_{ij} = \sigma \left(\delta_{ij}\frac{\partial C}{\partial
x_{k}}\frac{\partial C}{\partial x_{k}} - \frac{\partial C}{\partial
x_{i}}\frac{\partial C}{\partial x_{j}}\right), \quad i,j,k = 1, 2
\end{equation}
where $\delta_{ij}$ is the Kronecker delta. One important property
of the surface stress is that one of the principle axes $x'_1$ is
aligned with the gradient of the color function, the other principle
axis $x'_2$ is aligned with the interface tangential direction,
along which the only non-zero component of the surface stress is the
positive normal stress (tension), $\Pi_{2'2'} = \sigma |\nabla
C|^2$. The relation between the surface tension $\gamma$ and
$\sigma$ for an infinite plane interface is given by
\begin{equation}\label{surface-tension}
\gamma = \sigma \int^{+\infty}_{-\infty} \Pi_{2'2'} dx'_1.
\end{equation}

\begin{figure}[p]
\begin{center}
\includegraphics[width=\textwidth]{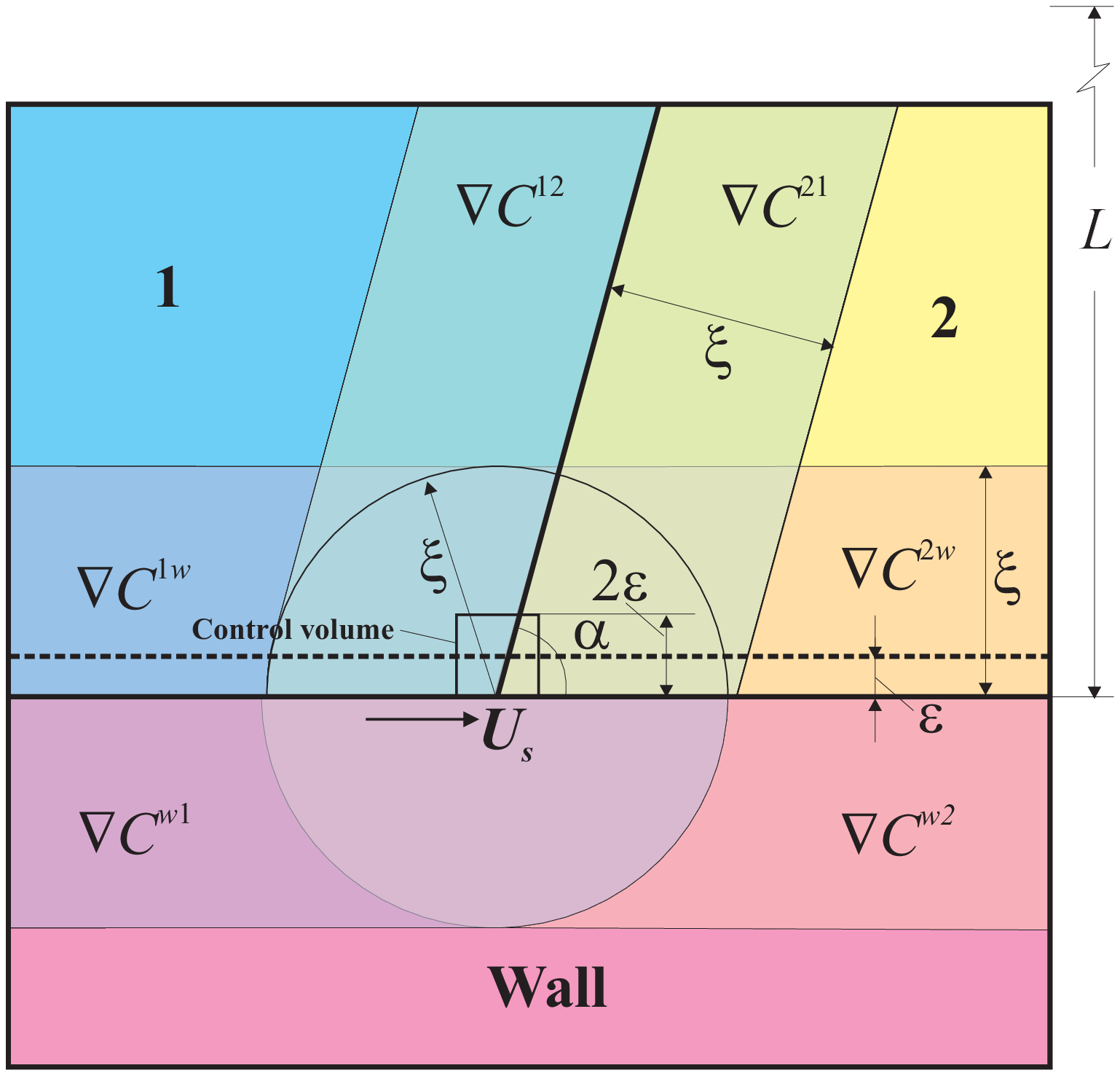}
\end{center}
\caption[]{smoothed, finite width interface model for moving contact
line problem}\label{triple}
\end{figure}

We consider a steady moving contact line with dynamic contact angle
$\alpha$, and velocity $U_s$, as shown in Fig. \ref{triple}.
Around the contact line there are three phases: fluid 1, fluid 2 and
the static wall. We define the color function as
\begin{equation}\label{color}
 C^{kl} = \begin{cases}
  1  \quad \text{in phase $l$}\\
  0  \quad \text{else}
\end{cases}, \quad k, l = 1, 2, w.
\end{equation}
Note that the color function is discontinuous across the interfaces.
In order to obtain a finite, continuous surface stress, we introduce
a two-dimensional smoothing-kernel function \cite{Natanson1960}
\cite{Monaghan1992}
\begin{equation}\label{smoothing-function}
W(\mathbf{x, \xi}) = \frac{1}{\xi^2\pi}e^{- \mathbf{x}^2/\xi^2}
\end{equation}
in which $\xi$ is smoothing length and $\xi<< L$, $L$ is the
characteristic length scale of the system. $W(\mathbf{x, \xi})$ is
radially symmetric and has the properties $\int
W(\mathbf{x}-\mathbf{x}', \xi) d\mathbf{x} =1 $ and
$\lim_{\xi\rightarrow 0}W(\mathbf{x}-\mathbf{x}', \xi) =
\delta(\mathbf{x}-\mathbf{x}')$. After convolution with the kernel
function, the smoothed gradient of the color function pointing
towards phase $l$ at a point $\mathbf{x}'$ in phase $k$ is
\begin{equation}\label{smoothing}
\nabla C^{kl}(\mathbf{x}')  = \int C^{kl}(\mathbf{x}) \nabla
W(\mathbf{x}-\mathbf{x}', \xi) d \mathbf{x}, l \neq k.
\end{equation}
Assuming that the smoothed profile defined by $\nabla
C^{kl}(\mathbf{x}')$ is the interface profile corresponding to an
equilibrium form of the bulk energy density, the total surface
stress at a point in phase $k$ is can be calculated from Eq.
(\ref{surface-stress}), by $\Pi_{ij} = \sum_{l\neq k} \Pi_{ij}
(\nabla C^{kl})$. It is easy to verify by Eq.
(\ref{surface-tension}) that, for a infinite plane interface between
phase $k$ and phase $l$, the surface tension is
\begin{equation}\label{surface-tension-1}
\gamma_{kl} = \frac{\sigma_{kl}}{\sqrt{2\pi}\xi}.
\end{equation}
Figure \ref{triple} indicates the regions of non-vanishing $\nabla
C^{kl}$ for different phase pairings. Note that there are overlap
regions near the contact line.

In order to study the contact-line dynamics, as shown in Fig.
\ref{triple}, we define a small square control volume with side
length $2\varepsilon$, $\varepsilon \ll \xi$, with one side on the
wall surface so that the interface between the fluid 1 and fluid 2
goes through the control volume. Assuming incompressibility and
straight interfaces with $\alpha$ not far from $\frac{\pi}{2}$, we
obtain
\begin{equation}\label{balance-x}
2\varepsilon \Pi^{l}_{11}  + \int \Pi^{w}_{21}dx_1 + 2\varepsilon
\tau^{l}_{11} + \int \tau^{w}_{21} dx_1= 2\varepsilon \Pi^{r}_{11}
 + \int \Pi^{f}_{21} dx_1 + 2\varepsilon \tau^{r}_{11} +
\int\tau^{f}_{21} dx_1 ,
\end{equation}
by considering the force balance on the control volume in tangential
(wall parallel) direction, where the superscripts $l$, $w$, $r$ and
$f$ represent the left, wall, right and upper faces of the control
volume. $\Pi_{11}$ and $\Pi_{21}$ are the tangential components of
surface stress. $\tau_{11}$ and $\tau_{21}$ are the tangential
components of viscous stress. As $\varepsilon$ is small and
$\varepsilon \ll \xi$ , the gradients of the color functions at a
point on the face of control volumes can be approximated with the
representative values on the contact line. If the contact line is
defined to be at the origin of a two-dimensional polar coordinate
system, the gradient is given by
\begin{equation}\label{gradient}
\nabla C^{kl}(\mathbf{x}'\rightarrow 0) = \int_{\Omega(\theta,
\theta')} \nabla W(\mathbf{x}, \xi)d \mathbf{x}, l \neq k
\end{equation}
where $\Omega(\theta, \theta')$ represents the sector between polar
angles $\theta$ and $\theta'$ in two-dimensional polar coordinates,
and $\theta, \theta' = 0, \alpha, \pi$ depending on the choice of
phase pairs. It can be readily obtained that $\int_{\Omega(\theta,
\theta')} \nabla W(\mathbf{x}, \xi) d\mathbf{x} = (\sin \theta -
\sin \theta', \cos \theta' - \cos \theta)\frac{1}{2\sqrt{\pi}\xi}$.
With Eq. (\ref{surface-stress}), the tangential surface-stress
components in Eq. (\ref{balance-x}) are $\Pi^{12}_{11} =
\frac{\sigma_{12}}{4\pi\xi^2}(1-\cos \alpha)^2$, $\Pi^{12}_{21} =
\frac{\sigma_{12}}{4\pi\xi^2}\sin \alpha (\cos \alpha - 1)$,
$\Pi^{21}_{11} = \frac{\sigma_{12}}{4\pi\xi^2}(1+\cos \alpha)^2$,
$\Pi^{21}_{21} = \frac{\sigma_{12}}{4\pi\xi^2}\sin \alpha (1+\cos
\alpha)$, $\Pi^{1w}_{11} = \frac{\sigma_{1w}}{\pi\xi^2}$,
$\Pi^{1w}_{21} = 0$, $\Pi^{2w}_{11} = \frac{\sigma_{2w}}{\pi\xi^2}$
and $\Pi^{2w}_{21} = 0$, where $\sigma_{1w}$, $\sigma_{1w}$ and
$\sigma_{12}$ are the coefficients between fluid 1 and wall,  fluid
2 and wall, and fluid 1 and fluid 2.  Hence, using Eq.
(\ref{surface-tension-1}), Eq. (\ref{balance-x}) becomes
\begin{equation}\label{balance-x-result}
\gamma_{1w} + \sqrt{\frac{\pi}{2}}\xi\mu\left[\left(\frac{\partial
u_1}{\partial x_2}\right)^{w} - \left(\frac{\partial u_1}{\partial
x_2}\right)^{f} + \left(\frac{\partial u_1}{\partial x_1}\right)^{l}
- \left(\frac{\partial u_1}{\partial x_1}\right)^{r}\right] =
\gamma_{2w} + \frac{1}{2}\gamma_{12}\cos \alpha
\end{equation}
where $\gamma_{1w}$, $\gamma_{1w}$ and $\gamma_{12}$ are the surface
tensions between fluid 1 and wall, fluid 2 and wall, and fluid 1 and
fluid 2, respectively. Note that, the shear rates on the faces of
the control volume are approximated with the values on face-centers,
and that the viscosities of fluid 1 and fluid 2 are assumed to have
the same value $\mu$. When the fluids are in static equilibrium the
second term on the left-hand-side disappears, Eq.
(\ref{balance-x-result}) becomes
\begin{equation}\label{balance-static}
\gamma_{1w} = \gamma_{2w} + \frac{1}{2}\gamma_{12}\cos \alpha '
\end{equation}
where $\alpha'$ is the static contact angle. Eq.
(\ref{balance-static}) implies that the static contact angle is
different from that obtained by Young's relation, except $\alpha ' =
\frac{\pi}{2}$. This is not unexpected because the current relation
gives the force balance within the interface. Note that for $\alpha
\neq \alpha '$, an unbalanced surface stress along the tangential
direction arises in Eq. (\ref{balance-x}), and is balanced by the
differences between the shear stresses. To study the details of the
balance between surface forces and viscous forces, it is convenient
to define a layer, as shown in Fig. \ref{triple}, which has a small
thickness of $\varepsilon$ and a velocity $U_\varepsilon$ in the
center of the control volume. As $U_\varepsilon \rightarrow U_s$ for
$\varepsilon \rightarrow 0$, one can study the force balance exactly
at the contact line. For any location other than the contact line
there is no unbalanced surface stress as in Eq. (\ref{balance-x})
and hence viscous forces are continuous. Therefore, it is
straightforward to assume that the fluid velocity on the left and
right faces of the control volume match continuously with the wall
velocity, and the shear rate on the upper face of the control volume
match continuously with that of the bulk flow. Here, three types of
wall boundary conditions with different wall-slips, i.e. no-slip,
finite-slip and free-slip, are to be considered.
\section{Discussion}
If a no-slip wall boundary condition is applied, the viscous stress
is calculated from viscosity and shear rate. As $\varepsilon$ is
small, a linear approximation of the shear rates is sufficient, then
Eq. (\ref{balance-x-result}) can be rewritten as
\begin{equation}\label{shear-surface-balance}
\Gamma \mu U_\varepsilon - \sqrt{2 \pi} \xi \mu \left(\frac{\partial
u_1}{\partial x_2}\right)^{f} = \sigma_{12}(\cos \alpha - \cos
\alpha').
\end{equation}
where $\Gamma = \sqrt{2\pi}\frac{\xi}{\varepsilon} \gg 1$ is a
non-dimensional parameter. Note that the normal viscous stresses
here cancel out because of opposite directions and same magnitudes.
For the distinguished limit of a sharp interface $\Gamma \gg 1$ as
$\xi$, $\varepsilon \rightarrow 0$, one has
\begin{equation}\label{slip-velocity}
\Gamma {\rm Ca} =\cos \alpha - \cos \alpha'
\end{equation}
where the capillary number is defined by ${\rm Ca} = \mu
U_s/\gamma_{12}$, since $\left(\frac{\partial u_1}{\partial
x_2}\right)^{f} \sim \frac{U}{L}$ is finite and $U_\varepsilon
\rightarrow U_s$. Now that the same form as Eq.
(\ref{slip-velocity}) with $\Gamma = \sqrt{\frac{9 \pi}{2}}
\frac{\xi}{\varepsilon}$ can be derived if the viscosity of fluid 2
is neglected. Here, the problem discussed in Refs.
\cite{HuhScriven1971} and \cite{Dussan1979} can be solved: for
infinite viscous stress in Eq. (\ref{balance-x}) in the limit
$\varepsilon \rightarrow 0$, Eq. (\ref{slip-velocity}) implies that
there still is a contact-angle condition at the contact line for a
non-vanishing contact-line velocity $U_s$. The reason is that, for
$\xi \rightarrow 0$ the unbalanced surface stress in Eq.
(\ref{balance-x}) becomes infinite as well, and Eq.
(\ref{slip-velocity}) gives the condition for which the two infinite
stresses are in equilibrium. Note that the problem of a singular
viscous force remains only if $\varepsilon \rightarrow 0$ and $\xi$
does not vanish. This limit results in $U_s \rightarrow 0$, i.e. the
contact line does not move.

A straightforward interpretation of Eq. (\ref{slip-velocity}) with
respect to the microscopic length scales indicates that $\xi$
corresponds to the physical width of the interface and $\varepsilon$
to the thickness of the first molecules layer at the wall surface,
and $\Gamma$ is just the ratio between the two length scales. Since
the thickness of the first molecule layer is quite close to the
molecule size and the interface width is at least several times the
molecule size, our result $\Gamma \gg 1$ is physically meaningful.
Eq. (\ref{slip-velocity}) is derived from a classical hydrodynamic
analysis in which the only considered dissipation mechanism is the
viscous force. It is quite surprising that Eq. (\ref{slip-velocity})
has the same form as a linearized formulation of the
molecular-kinetic model \cite{Blake1993} which was proposed to
discard dissipation due to viscous flow: $\Gamma' {\rm Ca} = (\cos
\alpha - \cos \alpha')$, where $\Gamma'=\eta/\mu$, $\eta$ is the
coefficient of wetting-line friction. Note that $\eta$ has the units
of the viscosity $\mu$, and is always much larger than $\mu$
\cite{Blake2006}, which is in agreement with our result $\Gamma \gg
1$. It is also interesting that Eq. (\ref{slip-velocity}) has the
same form as the small-velocity-approximation relation of
Shihkmuraev's interface relaxation model \cite{Shihkmuraev1997}.
However, Shihkmuraev obtained a zero contact-line velocity for
negligible interface relaxation, as opposite to finite contact-line
velocity obtained in our current analysis.

If a finite-slip wall boundary condition is applied, the viscous
stress on the wall for $\varepsilon \rightarrow 0$ is given as
$\beta U_s$, where $\beta$ is the slip-coefficient. In this case Eq.
(\ref{balance-x-result}) becomes
\begin{equation}\label{shear-surface-balance-finite}
\beta U_s - \mu \left(\frac{\partial u_1}{\partial x_2}\right)^{f} =
\sqrt{2 \pi}\frac{\sigma_{12}}{\xi}(\cos \alpha - \cos \alpha').
\end{equation}
Eq. (\ref{shear-surface-balance-finite}) is in agreement with the
slip-wall boundary condition of Qian, Wang \& Sheng
\cite{QianWangSheng2003}, which states that the wall slip is
proportional to the sum of the viscous stress and the uncompensated
Young stress. An important result different from that of the sharp
interface model is that for a given contact-line velocity the
dynamic contact angle is strongly affected by the shear rate of the
bulk flow. Note that Eq. (\ref{shear-surface-balance-finite}) is
valid only for finite-thickness interfaces and implies a
surface-force singularity if the interface thickness tends to zero.
To eliminate the singularity the dynamic and static contact angles
should be equal $\alpha = \alpha'$, which explains the underlying
reason of Cox's hypothesis \cite{Cox1986} for a macroscopic analysis
stating that wall-slip is permitted and the contact angle is
independent of contact-line velocity. A result of the surface-force
balance with $\alpha = \alpha'$ from Eq.
(\ref{shear-surface-balance-finite}) is the widely used Navier
boundary condition $U_s = \lambda \left(\frac{\partial u_1}{\partial
x_2}\right)^{f}$, where $\lambda = \mu /\beta$ is the slip length.

Our analysis does not allow for a free-slip boundary condition
\cite{HuhMason1977} because free slip would result in the first term
on the left-hand-side of Eq. (\ref{shear-surface-balance-finite}) to
vanish, which may lead to an un-physical an decrease of the contact
angle for an advancing contact line. For the sharp-interface limit
with free slip there is no force balance, no matter whether the
dynamic and static contact angles are the same or not.
\section{Conclusion}
To summarize, we have studied the force balance at a moving contact
line with different boundary conditions. It is found that, for the
sharp-interface limit, both the finite-slip and the no-slip wall
boundary conditions are possible. With the finite-slip wall
assumption, the analysis explains that the previously used Cox's
hypothesis and the Navier boundary condition are essential for a
force balance. However, the analysis also suggests that the no-slip
wall boundary conditions is still valid along with a contact-angle
condition, which agrees with several previous studies on dynamic
contact angles. More importantly, since there is no conflict with
the results of molecular dynamics simulations, the no-slip wall
boundary condition can serve for obtaining more reliable numerical
predictions of moving contact line problems.
\bibliographystyle{plain}
\end{document}